\documentclass[12pt]{article}
\RequirePackage[colorlinks,citecolor=blue,urlcolor=blue]{hyperref}

\usepackage{amssymb}
\usepackage{amsfonts}
\usepackage{amsmath}
\usepackage{hyperref}

\usepackage{epsf}
\usepackage{epsfig}
\usepackage{graphicx}
\usepackage{caption}
\usepackage{subcaption}

\newcommand{\E}{\mathbb E}
\newcommand{\Q}{\mathbb Q}

\newcommand{\be}{\begin{equation}}
\newcommand{\ee}{\end{equation}}

\usepackage{graphicx}
\graphicspath{{converted_graphics/}}

\begin{document}

\title{The Generalized Gamma distribution as a useful RND under Heston's stochastic volatility model}
\author{Ben Boukai \\
Department of Mathematical Sciences, IUPUI \\
Indianapolis, IN 46202 , USA}
\maketitle

\begin{abstract}

\noindent Following Boukai (2021) we present the Generalized Gamma (GG) distribution as a possible  RND for modeling European options prices under Heston's (1993) stochastic volatility (SV) model. This distribution is seen as especially useful in situations in which the spot's price follows a negatively skewed distribution and hence, Black-Scholes based (i.e. the log-normal distribution) modeling is largely inapt.  We apply the GG distribution as RND to modeling current market option data on three large market-index ETFs, namely the {\tt SPY, IWM} and {\tt QQQ} as well as on the {\tt TLT} (an ETF that tracks an index of long term US Treasury bonds).  The current option chain of each of the three market-index ETFs shows of a pronounced skew of their volatility `smile' which indicates  a likely distortion  in the Black-Scholes modeling of such option data. Reflective of entirely different market expectations, this distortion appears not to exist in the {\tt TLT} option data. We provide a thorough modeling of the available option data we have on each ETF (with the October 15, 2021  expiration) based on the GG distribution and compared it to the option pricing  and RND modeling obtained directly from a well-calibrated Heston's (1993) SV model (both theoretically and empirically, using Monte-Carlo simulations of the spot's price).   All three market-index ETFs exhibit negatively skewed distributions which are well-matched with those derived under the GG distribution as RND. The inadequacy of the Black-Scholes modeling in such instances which involve negatively  skewed distribution  is further illustrated by its impact on the hedging factor, delta,  and the immediate implications to the retail trader. In contrast, for the {\tt TLT} ETF, which exhibits no such distortion to the volatility `smile',  the three pricing models (i.e. Heston's, Black-Scholes and Generalized Gamma) appear to yield similar results. 

\bigskip

\end{abstract}

\section{Introduction }
One of the most widely celebrated option pricing model for equities (and beyond) is that of Black and Scholes (1973) and of Merton (1973), (abbreviated here as the BSM).  Their option pricing model   was derived under some simple assumptions concerning the distribution of the asset's returns, coupled with presumptive continuous hedging, self-financing, zero dividend, risk-free interest rate, $r$,  and no cost of carry or transactions fees. In its standard form, the BSM model assumes that the spot's price process ${\cal{S}}=\{S_t, \, t\geq 0\}$  evolves with a constant volatility of the spot's returns, $\sigma$,   as a geometric Brownian motion (under a risk-neutral probability measure $\Q$, say), leading to  an exact solution for the price, $C(\cdot)$,  of an European call option. 
Specifically, given the {\it current} spot price $S_\tau=S$ and the risk-free interest rate $r$, the price of the corresponding call option with price-strike $K$ and duration $T$, 
\be\label{1}
C_{S}(K)= S\, \Phi(d_1)-K\,  e^{-rt}\, \Phi(d_2),  
\ee
where $t=T-\tau$ is the {\it remaining} time to expiry. Here, we use  the conventional notation to denote by  $\Phi(\cdot)$ and $\phi(\cdot)$ the standard Normal cumulative distribution function ($cdf$) and density function ($pdf$), respectively,  and where
\be\label{2}
d_1:=\frac{-\log(\frac{K}{S})+(r+\frac{\sigma^2}{2})t}{\sigma\sqrt{t}} \qquad \text{and} \qquad d_2:=d_1-\sigma\sqrt{t}.  
\ee
Despite of its wide acceptability in the retail trading world\footnote{Nowadays, many of the retail brokerage houses  operate entirely within the ``Black-Scholes world'' and provide, aside from market option bid-ask prices, the related ``Greeks'' and implied volatility values derived from and calculated under the BSM formula (\ref{1})-(\ref{2})} this model hinges on several incorrect assumptions and  hence, suffers from some notable deficiencies; see for example Black (1988, 1989) who pointed out {\it{the holes in Black-Scholes}}. Chief among the noted deficiencies is the fact that volatility of spot's returns (i.e., $\sigma$) appears not to be constant over the  \textit{`life'} of of the option, but rather varying at random. 

The efforts to incorporate a non-constant volatility in the option valuation  (e.g., Wiggins (1987) or Stein and Stein (1991)) has culminated with Heston's (1993) stochastic volatility (SV) model.  This SV model incorporates,  aside from the dynamics of the spot's  price process ${\cal{S}}$, also the  dynamics of  a corresponding, though unobservable (hence untradable),  volatility process ${\cal{V}}=\{V_t, \, t\geq 0\}$.  Instructed by the {\it form} of the  exact BSM solution in (\ref{1}), Heston (1993)  obtained  that the solution to the  system of PDE he obtained from the stochastic volatility model he constructed is given by
\be\label{3}
C_{S}(K)= S\,  P_1-K\,  e^{-rt}\,  P_2,  
\ee
where $P_j$ $j=1,2$, are two related (under  $\Q$)  conditional probabilities that the option will expire in-the-money, conditional on the given current stock price $S_\tau=S$ and the current volatility, $V_\tau=V_0$.  However, unlike the explicit BSM solution in (\ref{1}) which is given in terms of the normal (or log-normal) distribution, Heston (1993) provided  (semi)  closed-form solutions to these two probabilities, $P_1$ and $P_2$,  both given in terms of their characteristic functions. These characteristic functions depend on some parameters of the SV model, $\vartheta=(\kappa, \theta, \eta, \rho)$ and may be numerically evaluated, via complex integration,  for any choice of the parameters $\vartheta$  in addition to the given $S, \ V_0$ and $r$ (for more details, see the Appendix).  The components of $\vartheta$ have  particular meaning in context of Heston's (1993)  SV  model: $\rho$ is the correlation between the random components of the spot's price and volatility processes, ; $\theta$ is the long-run average volatility, $\kappa$ is the mean-reversion speed for the volatility dynamics and $\eta^2$ is the variance of the volatility $V$. It should be noted that different choices of $\vartheta$ will lead to different {\it values} $C_{S}(K)$ in (\ref{3}) and hence, the value ${\vartheta}=(\kappa, \theta, \eta, \rho)$  must be appropriately {\it `calibrated'} first for $C_{S}(K)$ to actually match the option market data. 
However, this calibration process typically involves substantial numerical challenges (resulting from the complex-domain integration and the required multi-dimensional optimization)  and in light of its obvious complexity is not readily available to the retail option trader for the evaluation of $C_{S}(K)$ in (\ref{3}). 

 On the other hand, as was  established by Cox and Ross (1976), the risk-neutral option valuation (under $\Q$) provides that 
 that for $T>\tau$ (with $t=T-\tau$),  $C_{S}(K)$ must also satisfy

\be\label{4}
C_{S}(K)=  e^{-rt} \int_K^\infty (S_T-K)\,   q(S_T)dS_T,
\ee
where $q(\cdot)$ is the density of some  risk-neutral distribution (RND) $Q(\cdot)$, under the probability $\Q$, reflective of the conditional distribution of the spot price $S_T$ at time $T$, given the spot price, $S_\tau$ at time $\tau<T$ whose expected value is the future value of the spot's price. Namely, the RND $q(\cdot)$ must also satisfy, 
\be\label{5}
\E (S_T| S_\tau=S) =    \int S_T\cdot q(S_T)d S_T= S \cdot e^{rt}. 
\ee
This risk-neutral density  (or distribution)  links together (under   $\Q$) for the option evaluation the distribution of the spot's price $S_T$ and the stochastic dynamics governing the underlying model. As was mentioned earlier, in the case of the BSM in (\ref{2}) the RND is unique and is given by the log-normal distribution. However, since the Heston (1993) model involves the dynamics of two stochastic processes, one of which (the volatility, ${\cal{V}}$) is untradable and hence not directly observable, there are innumerable many choices of RNDs,  $q(\cdot)$, that would satisfy (\ref{4})-(\ref{5}) and hence, the general solutions of $P_1$ an $P_2$ in (\ref{3}) by means of characteristic functions (per each choice of the structural parameter $\vartheta=( \kappa, \theta, \eta, \rho)$).

\subsection{The Heston's RND as a class of scale-parameter  distributions}

In the literature, one can find numerous papers dealing with the {\it `extraction', `recovery', `estimation'} or {\it  `approximation'}, in parametric or non-parametric frameworks,  of the RND, $q(\cdot)$ from the available (market) option prices. Some comprehensive literature reviews of the subject can be found in Jackwerth (2004), Figlewski (2010), Grith and Krätschmer  (2012)  and  Figlewski (2018). In particular, within the parametric approach, one attempts  to estimate by various standard means (maximum likelihood, method of moments, least squares, etc.)  the parameters of some {\it assumed}  distribution so as to approximate  available option data or implied volatilities (c.f. Jackwerth and Rubinstein (1996)). This type of {\it assumed} multi-parameter distributions includes some mixtures of log-normal distribution (Mizrach (2010), Grith and Krätschmer (2012)),  generalized gamma (Grith and Krätschmer  (2012)), generalized extreme value (Figlewski (2010)), the gamma and the Weibull distributions (Savickas (2005)), among  others. While empirical considerations have often led to suggesting these particular parametric distributions as possible $pdf$ in (\ref{4}), the motivation to these considerations did not include direct link to the governing pricing model and its dynamics, as was the case in the  BSM model, linking directly the log-normal distribution and the price dynamics reflected by the geometric Brownian motion leading to the BSM formula in (\ref{1}).

In a recent paper,  Boukai (2021) has identified the class of distributions that could serve as RNDs for Heston's (1993)  SV model. His  theoretical result   is built on the realization that any RND $Q(\cdot)$  in (\ref{4}) that satisfies Heston's (1993) model and hence (\ref{3}) may be presented as
\be\label{6}
C_{S}(K) \equiv  S\cdot \Delta(K)-K\,  e^{-rt}\, \cdot (1-Q(K)), 
\ee
where  $\Delta(K)$  denotes the so-call {\it delta} function (or hedging fraction) in the option valuation, as defined by
\be\label{7}
\Delta(K)=\frac{\partial C_{S}(K)}{\partial S}. 
\ee
In particular, it was shown that the class of scale-parameter distributions with mean being the
forward spot's price, $\mu:=S\cdot e^{rt}$  would admit the presentation in (\ref{6}) and hence would satisfy Hestons' (1993) option pricing model in (\ref{3}). In fact, it was also shown that the RNDs that may be calculated directly from Heston's characteristic functions (corresponding to $P_1$ and $P_2$) are members of this class of distributions as well.  Accordingly, Boukai's (2021) main results (as are summarized in Theorem 2 there) establish the direct link, through Heston's (1993) solution in (\ref{3}) (or  (\ref{6})) between this class of RNDs and the assumed stochastic volatility model   governing the spot price dynamics.  
 To fix ideas, we set $\mu=S\cdot e^{rt}$ to denote the forward spot's price and we correspondingly denote by $Q_\mu(\cdot)$  the  RND with a  corresponding $pdf$ $q_\mu(\cdot)$ as in  (\ref{4}) - (\ref{5}).  It is assumed that $\mu$ is a scale parameter of  $Q_\mu(\cdot)$  so that for any $x>0$,  $Q_\mu(x)\equiv Q_1(x/\mu)$ and $q_\mu(x)\equiv q_1(x/\mu)/\mu$  for \emph{some} $cdf$ $Q_1(\cdot)$ with a $pdf$ $q_1(\cdot)$ satisfying   $\int_0^\infty x q_1(x)dx=1$ and  $ \int_0^\infty x^2 q_1(x)dx=1+\nu^2$. Here $\nu^2$ is some exogenous parameter (to be specified later). In Theorem 1 of Boukai (2021) it was shown that for this class of scale-parameter distributions,  the Delta function  (\ref{7}) may in fact be written as
 \be\label{8}
\Delta(K)\equiv \Delta_\mu(K)=\frac{1}{\mu} \int_{K}^\infty xq_\mu(x)dx\equiv \Delta_1(K/\mu), 
\ee
 where  $\Delta_1(s):=\int_s^\infty uq_1(u)du$ for any $s>0$.  Accordingly,  for any member of this class of scale-parameter (in $\mu=S\cdot e^{rt}$), $C_S(K)$ in (\ref{6}) may equivalently be written as 
\be\label{9}
C_{S}(K) \equiv  S\cdot \Delta_1(K/\mu)-K\,  e^{-rt}\, \cdot (1-Q_1(K/\mu)). 
\ee
 Thus,  any member of this scale-parameter class of distributions defined by $Q_1(\cdot)$ could be used for the direct risk-neutral valuation of the option price under Heston's SV model. The expression in (\ref{9}) is our `working' formula for the direct calculations of the option price $C_S(K)$ in the case of scale-parameter distribution defined by $Q_1(\cdot)$.   This result was illustrated in great details  by Boukai (2021) for one-parameter versions of the log-Normal (i.e. the BSM model), Inverse-Gaussian, Gamma, Inverse-Gamma, Weibull and the Inverse-Weibull distributions which provide explicit RNDs for Heston's pricing model in various market circumstance (e.g., negatively skewed RND to match {\tt{SPX}} option data, or positively skewed RND to match {\tt{AMD}} option data).

\subsection{An Overview}

In this paper we focus attention on a two-parameter version of the Generalized Gamma (GG) distribution as is especially parametrized to serve  as a RND under the Heston's SV option valuation model. The particular version of this distribution we consider here is characterized  by two shape parameters $\alpha$ and $\xi$ say,  and is general enough to admit either positively skewed distributions ($\xi<0$) or negatively skewed distributions $(\xi>0)$.  Aside from this noted `elasticity' to match well varying characteristics of different spot's RNDs,  this distribution is especially useful in modeling option prices in situations that exhibit put-over-call skew and and hence admit negatively skewed distribution of the spot's price.  In Section 2 we present the Generalized Gamma distribution and reparametrize it so that it may serve as a   RND under the Heston (1993) Stochastic Volatility model. Though not of immediate interest we also present in Section 2.2 the Inverse Generalized Gamma (IGG) distribution as a possible RND under Heston's SV model that could be useful  in modeling positively skewed (implied) distributions.  

 we apply the GG distribution as a RND to modeling current market option prices on three large market ETFs, the {\tt SPY, IWM} and {\tt QQQ}.  The current option chain for these three ETFs a pronounced skew of their volatility `smile' which indicates of a likely distortion  in the Black-Scholes modeling of such option data. We provide a thorough modeling of the available option data we have on each ETF (the October 15, 2021 with 63 to expiration) based on the Generalized Gamma Distribution and compared it to the option pricing  and RND modeling obtained directly from a well-calibrated Heston's (1993) SV model (both theoretically and empirically, using Monte-Carlo simulations of the spot's price).   All three ETFs exhibit negatively skewed distributions which are well-matched with those derived from the Generalized Gamma Distribution. The inadequacy of the Black-Scholes modeling in such instances with negatively  skewed distribution  is further illustrated by its impact on the hedging factor, delta and the immediate implication to the retail trader. Some details on Heston's SV model and characteristic functions are provided in the  Appendix. 

 In Section 3 we apply the GG distribution as RND to modeling current market option data on three large market-index ETFs, namely the {\tt SPY, IWM} and {\tt QQQ} as well as on the {\tt TLT} (an large ETF that tracks an index of long term US Treasury bonds).  The current option chain of each of the three market ETFs shows of a pronounced skew of their volatility `smile' which indicates  a likely distortion  in the Black-Scholes modeling of such option data. Reflective of entirely different market expectations, this distortion appears not to exist in the {\tt TLT} option data (see Figure 1 below). We provide a thorough modeling of the available option data we have on each ETF (with the October 15, 2021  expiration) based on the GG distribution and compared it to the option pricing  and RND modeling obtained directly from a well-calibrated Heston's (1993) SV model (both theoretically and empirically, using Monte-Carlo simulations of the spot's price).   All three market-index ETFs exhibit negatively skewed distributions which are well-matched with those derived under the GG distribution as RND. The inadequacy of the Black-Scholes modeling in such instances which involve negatively  skewed distribution  is further illustrated by its impact on the hedging factor, delta,  and the immediate implications to the retail trader. In contrast, for the {\tt TLT} ETF, which exhibits no such distortion to the volatility `smile',  the three pricing models (i.e. Heston's, Black-Scholes and Generalized Gamma) appear to yield very similar results. Technical notes are provided in Section 3.2 and some details on Heston's SV model and related characteristic functions are provided in the  Appendix.

\section{The Generalized Gamma distribution as a Heston's RND}
   
Introduced by Stacy (1962), the Generalized Gamma  (GG) distribution  is demonstrably highly versatile, with a vast number applications,  from survival analysis to meteorology and beyond (see for example  Kiche, Ngesa and Orwa (2019), Thurai and Bringi (2018).  It includes among many others, the Weibull distribution, the Gamma distribution  and the log-normal distribution as a limiting case.  In this section we show that this distribution and its counterpart, the so-called Inverse Generalized Gamma distribution (IGG), both under a particular re-parametrization,  satisfy the conditions of Theorem 2 in Boukai (2021) and hence, could serve as  RND (for direct option valuation using (\ref{9})) under Heston's (1993) stochastic volatility  model for option valuation.  Though similar, we will present  these two cases of the generalized gamma distribution separately (as in the Weibull case discussed in Examples 34 and 3.5 of Boukai (2021))  as  they do present different profiles of skewness and kurtosis.  We will however focus our attention on the GG distribution, as we will use for option pricing modeling in situation which involved negatively skewed (implied) risk-neutral distributions.  
 
 We begin with some standard notations.  We write $Y\sim {\cal G}(\alpha, \lambda)$ to indicate that the random variable $Y$ has the 
gamma distribution with a scale parameter $\lambda>0$ and a shape parameter $\alpha>0$, (so that its mean is $\E(Y)=\alpha/\lambda$).  We  write $g(\cdot; \alpha, \lambda)$ and $G(\cdot; \alpha, \lambda)$ for the  corresponding $pdf$ and $cdf$ of $Y$, respectively, 
\be\label{10}
g(y; \alpha, \lambda)\equiv \frac{\lambda^\alpha y^{\alpha-1 }e^{-\lambda y}}{\Gamma(\alpha)} \qquad  \text{and} \qquad G(y; \alpha, \lambda)\equiv \frac{\Gamma(y\lambda; \alpha)}{\Gamma(\alpha)}, 
\ee
 where $\Gamma(\alpha):=\int_{0}^\infty x^{\alpha-1} e^{-x}dx$  denotes the  gamma function whose incomplete version is $\Gamma(s;  \ \alpha):=\int_{0}^s x^{\alpha-1} e^{-x}dx$, is defined for any $s>0$.   

\subsection{The GG Distribution} 

The Generalized Gamma (GG) distribution is typically characterized  by three parameters: a scale parameter, $\lambda>0$, and two shape parameters, $ \alpha>0$ and $\xi>0$ and is defined as follows. We say that  $W\sim {\cal GG}(\lambda,  \xi, \alpha)$, if 
\be\label{11}
Y\equiv  \left(\frac{W}{\lambda}\right)^\xi\sim  {\cal G}(\alpha, 1).
\ee
 With some additional restrictions on $\xi$, one can similarly define the so-called Inverse Generalized distribution (IGG). Namely, we say that  $W\sim {\cal IGG}(\lambda,  \xi, \alpha)$, if 
\be\label{12}
Y\equiv  \left(\frac{W}{\lambda}\right)^{-\xi}\sim  {\cal G}(\alpha, 1).
\ee

 In light of relation (\ref{11}),  the $cdf$  and $pdf$  of $W\sim {\cal GG}(\lambda,  \xi, \alpha)$,  are readily available in terms of the Gamma distribution in (\ref{10}). More specifically, for any $w>0$, 
 $$
 F_W(w):=Pr(W\leq w)= G\left((\frac{w}{\lambda})^\xi\ ; \alpha, \ 1\right), 
 $$
 and 
 $$
 f_W(w)= \frac{\xi}{\lambda} (\frac{w}{\lambda})^{\xi-1} \cdot g\left((\frac{w}{\lambda})^\xi\ ; \alpha, \ 1\right). 
 $$
 Also, the $j^{th}$ moment of this distribution (whenever exists, i.e. whenever $\alpha+j/\xi>0$ with $j=0, 1, 2,\dots$) is given by
 $ \E(W^j) =\lambda^j\cdot h_j(\xi)$, where for a given $\alpha>0$, $h_j(\xi):= \Gamma(\alpha+j/\xi)/{\Gamma(\alpha)}$ for $j=0, 1, \dots, $. 

Now suppose that for a given $\alpha>0$ a random variable $U$ has the 'standardized'  GG  distribution, with mean $\E(U)=1$ and a variance $Var(U)=\nu^2$, for some $\nu>0$,  (in fact, we will later take $\nu= \sigma\sqrt{t}$ for some $\sigma>0$). That is, for a given $\alpha>0$ and $\nu>0$, we let  $\xi^*\equiv \xi(\nu)$ be the (unique) solution of the equation 
\be\label{13}
\frac{h_2(\xi)}{h_1^2(\xi)}=1+\nu^2,
\ee
in which case, $h_j^*\equiv h_j(\xi^*), \ j=1, 2$, $\lambda^*\equiv 1/h_1^*$ and $U\sim {\cal GG}(\lambda^*, \xi^*, \alpha )$.  
Accordingly,  the $cdf$ of $U$ is given   by
\be\label{14}
Q_1(u):= Pr(U\leq u)= G\left((\frac{u}{\lambda^*})^{\xi^*}\ ; \alpha, \ 1\right), 
\ee
for any $u>0$, and its $pdf$ is given by
\be\label{15}
q_1(u):= \frac{\xi^*}{\lambda^*} (\frac{u}{\lambda^*})^{{\xi^*}-1} \cdot g\left((\frac{u}{\lambda^*})^{\xi^*}\ ; \alpha, \ 1\right), \ \ \ u>0.
 \ee

It  can be easily verified  that if $X\equiv \mu\cdot U$ for some $\mu>0$, then the $pdf$, $q_\mu(\cdot)$ of $X$ is the 'scaled' version of $q_1(\cdot)$ above. 
For this RND, the values of  $\Delta_1(s)$ in (\ref{8}) can be calculated by , for any $s>0$,  in a closed form as
\be\label{16}
\Delta_1(s)= \int_{s}^\infty uq_1(u)du =1-G((s/\lambda^*)^{\xi^*};\  \alpha+1/\xi^*, \ 1),
\ee
which, when combined in (\ref{9}) with the expression of $Q_1(\cdot)$, given in (\ref{14}) above,  provides the values of 
\be\label{17}
\begin{aligned}
c_\mu(k)=  &  \mu\times \left[\Delta_1({{k/\mu}})-\frac{k}{\mu}\times(1-Q_1(k/\mu))\right],\\
= & \mu \times \left[1-G((k/\mu\lambda^*)^{\xi^*};\  \alpha+1/\xi^*, \ 1)\right]- k \times \left[1-G((k/\mu\lambda^*)^{\xi^*};\  \alpha, \ 1)\right]\\
\end{aligned}
\ee
for any $\mu>0$.  Finally, to calculate under this generalized gamma  RND  the price of a call option  at a strike $K$ when the current price of the spot is $S$, we will utilize (\ref{17}) with $\mu\equiv S\, e^{rt}$, $k\equiv K$ and with $\lambda^*\equiv 1/h_1(\xi^*)$ and $\xi^*\equiv\xi(\nu)$ as is determined by equation (\ref{13}) above with $\nu\equiv  \sigma\sqrt{t}$ to obtain, $C_S(K)=e^{-rt}c_\mu(K)$ as,
\be\label{18}
C_S(K)=S\cdot \left[1-G(d;\  \alpha+1/\xi^*, \ 1)\right]- K e^{-rt} \cdot \left[1-G(d;\  \alpha, \ 1)\right], 
\ee
where 
$$
d=\left(\frac{ K e^{-rt}h_1(\xi^*) }{S}\right)^{\xi^*}, \qquad \text{with}\qquad \xi^*\equiv\xi(\nu) \ \ \text{ from (\ref{13})} . 
$$
We point out that for given current spot's price,  $S$, a strike price $K$,  risk free interest rate, $r$,  and the remaining option's duration $t$, the option value $C_S(K)$ in (\ref{18}) involves, through equation (\ref{13}) (with $\nu\equiv  \sigma\sqrt{t}$),   with only two parameters, namely $\alpha$ and $\sigma$. Their values can easily be {\it ``calibrated''} from the available market option data. Indeed in the Generalized Gamma case, this calibration task is computationally much simpler  than the direct  calibration of four parameters of Heston's pricing model,  based on the characteristic functions (Heston (1993); see Appendix)  which also involves integration over the complex domain.  This point is further demonstrated in Section 3 below.

 \subsection{The IGG Distribution} 
 For sake of completeness, we also present the details of this variant to the Generalized Gamma distribution here as well. With some additional restrictions on $\xi$, one can similarly define the Inverse Generalized Gamma distribution (IGG). Namely, we say that  $W\sim {\cal IGG}(\lambda,  \xi, \alpha)$, if 
\be\label{19}
Y\equiv  \left(\frac{W}{\lambda}\right)^{-\xi}\sim  {\cal G}(\alpha, 1). 
\ee
The option pricing model under the Inverse Generalized Gamma distribution as RND for the Heston's SV for option valuation, is constructed similarly to that the GG in the previous section.  By relation (\ref{19}), if  $W\sim {\cal IGG}(\lambda,  \xi, \alpha)$,  then its $cdf$ is given, for $w>0$,  
$$
 F_W(w):=Pr(W\leq w)= 1- G\left((\frac{w}{\lambda})^{-\xi}\ ; \alpha, \ 1\right). 
$$
 In this case too the 'standardized'  IGG  distribution of $U$, is constrained  to have mean $1$ and  variance $\nu^2$, which requires a restriction on the parameter $xi>2/\alpha$. It follows that with such a restriction, $U\sim {\cal IGG}(\lambda^*,  \xi^*, \alpha)$, but now,   $\xi^*\equiv \xi(\nu)$ is  the (unique) solution of the equation 
\be\label{20}
\frac{\tilde h_2(\xi)}{\tilde h_1^2(\xi)}=1+\nu^2,
\ee
$\tilde h_j(\xi)\equiv h_j(-\xi)= \Gamma(\alpha-j/\xi)/\Gamma(\alpha), \ j=1, 2. $, provided .  in which case, $\tilde h_j^*\equiv \tilde h_j(\xi^*), \ j=1, 2$, 
$\lambda^*\equiv 1/{\tilde h_1^*}$. Accordingly,  the $cdf$ of $U$ is given   by
\be\label{21}
Q_1(u):= Pr(U\leq u)=1- G\left((\frac{u}{\lambda^*})^{\xi^*}\ ; \alpha, \ 1\right), 
\ee
for any $u>0$, and in similarity to (\ref{16}), its corresponding delta function is given by
\be\label{22}
\Delta_1(s)=G((s/\lambda^*)^{-\xi^*};\  \alpha-1/\xi^*, \ 1),
\ee
Again, by combining (\ref{21}) and (\ref{22})  in (\ref{9}) we obtain that for any $\mu>0$, 
\be\label{23}
c_\mu(k)=  \mu \times G((k/\mu\lambda^*)^{-\xi^*};\  \alpha-1-\xi^*, \ 1)- k \times G((k/\mu\lambda^*)^{-\xi^*};\  \alpha, \ 1). 
\ee
Accordingly, in order to calculate under this Inverse Generalized Gamma RND  the price of a call option  at a strike $K$ when the current price of the spot is $S$, we will utilize (\ref{23}) with $\mu\equiv S\, e^{rt}$, $k\equiv K$ and with $\lambda^*\equiv 1/\tilde h_1(\xi^*)$ and $\xi^*\equiv\xi(\nu)$ as is determined by equation (\ref{20}) above with $\nu\equiv  \sigma\sqrt{t}$ to obtain, $C_S(K)=e^{-rt}c_\mu(K)$ as,
\be\label{24}
C_S(K)=S\cdot G(d;\  \alpha-1/\xi^*, \ 1)- K e^{-rt} \cdot G(d;\  \alpha, \ 1), 
\ee
where 
$$
d=\left(\frac{ K e^{-rt}\tilde h_1(\xi^*) }{S}\right)^{-\xi^*}, \qquad \text{with}\qquad \xi^*\equiv\xi(\nu) \ \ \text{ from (\ref{20})} . 
$$

\subsection{Skew and Kurtosis} 

As can be see from the above construction of the RNDs,  both the GG and IGG distributions depend on two shape parameters $(\alpha, \, \xi^*)$, or equivalently $(\alpha, \, \nu)$, where $\nu\equiv\sigma\sqrt{t}$, that affect their features, such as {\it kurtosis} and {\it skewness}, and hence their suitability as RND for various  particular scenarios of the SV model (\ref{40})-- as is determined by the structural model parameter  $\vartheta=(\kappa, \theta, \eta, \rho)$ (more on this point in the next section).   Unlike the standardized log-normal distribution which has a positive skew only, these two classes of distributions  offer a range of RNDs with positive as well as negative skewness.  This is a critical feature to have when modeling option prices for characteristically different spots such as an Index ( S\&P500 say) as oppose to modeling option prices for a technology firm (such as GME, say).  

For a given $(\alpha, \, \xi^*)$, we  denote these two measures as $\gamma_s(\xi^*)$ for {\it skew} and  
$\gamma_k(\xi^*)$  for the {\it kurtosis}. Then with $h_j(\xi):=\Gamma(\alpha+j/\xi)$ we have in the GG 
case that with $\xi^*=\xi(\nu)$ which satisfies (\ref{13}), 
$$
\gamma_1(\xi^*)=\frac{h_3(\xi^*)-3\nu^2-1}{\nu^3}
$$
and 
$$
\gamma_2(\xi^*)=\frac{h_4(\xi^*)-4\nu^3\gamma_1(\xi^*)-6\nu^2-1}{\nu^4}. 
$$
For the IGG case,  these two measure are similar and are given  by $\gamma_1(-\xi^*)$ and $\gamma_2(-\xi^*)$, provided that $\xi^*=\xi(\nu)$ as is determined by (\ref{20}), satisfies that  $\xi^*>4/\alpha$.

\section{Calibration, Validation and Examples}
\subsection{Observing the skew}

In this section we demonstrate  the usefulness of the Generalized Gamma distribution to serve as a RND under Heston's Stochastic Volatility model in cases that exhibit a high put - call skew (i.e., OTM puts in the option series are far more expensive  than equidistant OTM calls) and hence, expressing a pronounced skew in the so-called ``volatility  smile'' of the series.  As cases in point are traded market indexes such as the S\&P 500 ({\tt{SPX}}), Russel 2000 ({{\tt RUT}}) or Nasdaq 100 ({{\it NDX}}),  which all  are  (along with their corresponding ETF surrogates, {{\tt SPY, IWM}} and {{\tt QQQ}}) currently at (or near) their all time high levels\footnote{As of the writing of this paper, August 14, 2021}.  Market expectations of an eminent `correction'  are often seen as the culprits that affected the implied volatility surface associated with the corresponding option series of the index (see for example Bakshi, Cao and Chen (1997)).

{
\begin{figure}[t] 
 
  \includegraphics[width=6.5in,height=8.5in,keepaspectratio]{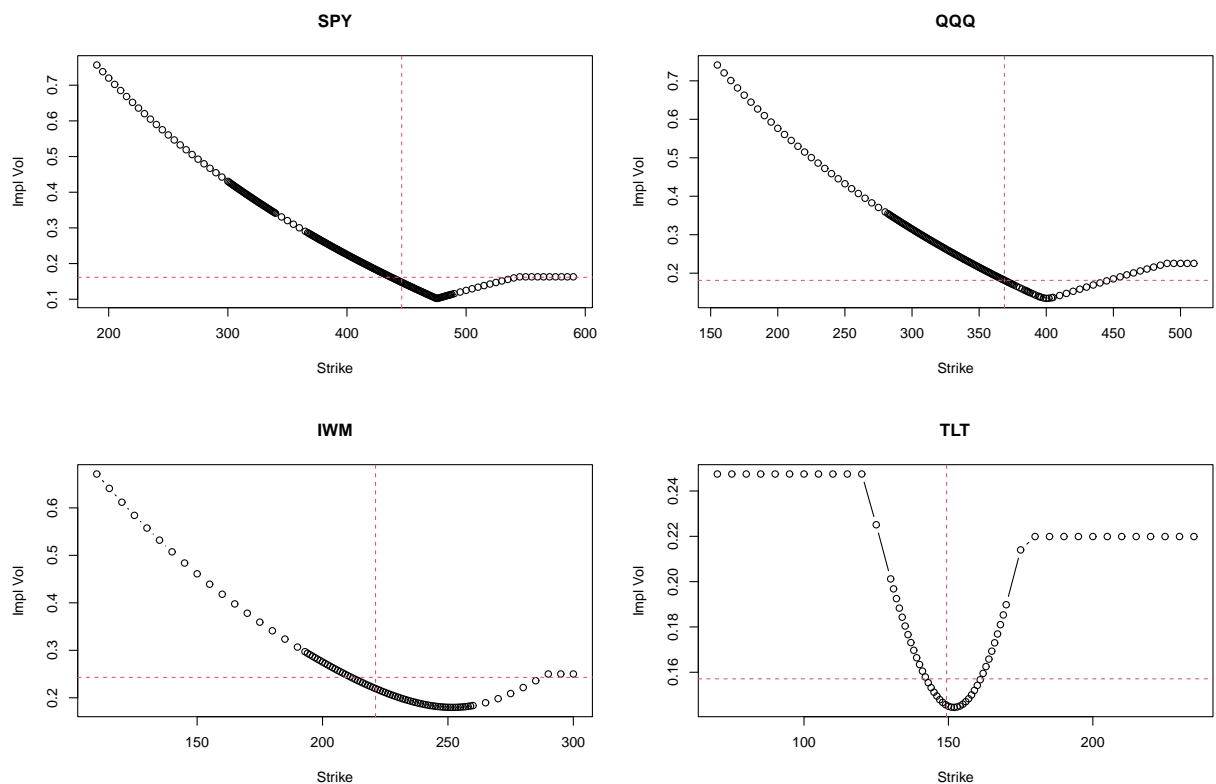}
 {\small{ \caption{\it The volatility `smiles' of the the October 15, 2021 option series (calls)  as observed and calculated on August 13, 2021 (EOD) for  the three market index ETFs, {{\tt SPY, IWM}} and {{\tt QQQ}} and on  August 18, 2021 (EOD)  for the {{\tt TLT}} ETF. }}}  
\end{figure}
}

Figure 1 above displays the calculated implied volatility smile of the October 15, 2021 option series for these three ETFs,  {{\tt SPY, IWM}} and {{\tt QQQ}}, as were quoted on August 13, 2021, each with 63 days to expiration (DTE).  Several  days later, on August 18, 2021, we obtained the corresponding quote for the {{\tt TLT}},  but now with 57 DTE.  For each ETF,  the EOD option's market prices (for puts and calls) at the corresponding  strikes were recorded along with the BSM-based calculated delta and implied volatility as were provided by the brokerage
 firm.\footnote{Option chain quotes were retrieved from TD Ameritrade  using the TOS platform} As reference, we also marked  on these plots (in red) the current spot's (ETF) price $S$ along with the ATM  (BS-based) calculated implied volatility (IV) for each ETF.    As can be seen from these figures, the options of the three market index ETFs exhibit a highly pronounced skew in their volatility `smile'  whereas the option on the {{\tt TLT}} ETF  do not (likely only reflective of market's expectations of actions by the Federal Reserve).

However, since typically  in the retail world, the calculated option's implied volatility (as well as other associated quantities, such as the option's delta) are all calculated based on the Black-Scholes formula in ({\ref{1}),  the noted distortion in the volatility smile (or surface) is nonetheless also indicative   that  the underlying log-normal distribution of the Black-Scholes model (with its distinctive positive skew measure) is a poor choice to serve as RND in such instances involving a stochastic volatility structure as that of Heston's (1993) (see (\ref{40}) below); particularly in those instances  which admit a negatively skewed RND. To illustrate the extent of the ``inaptness'' of using the log-normal distribution as RND  (the BS formula in (\ref{1})) for the option valuation in such skewed cases, we have {\it calibrated} for each of these four ETFs the  appropriate Heston's SV model to fit the observed market option data (i.e. on the October 15, 2021 option series for each) and derived from it the underlying RND of the Heston's model (HS). This RND which is obtained both theoretically, using (\ref{40.5}) and ({\ref{3}), and via Monte-Carlo simulations of (\ref{40}),  will serve us as a benchmark for comparison. 

 For each option series, the available market data consists of the $N$ strikes, $K_1, \dots, K_{N}$  with corresponding call option (market) prices $C_1, \dots, C_{N}$\footnote{These prices could be the actual market prices or the average between the bid and ask prices of the market}.  As standard measure of the {\it goodness-of-fit} between the model-calculated option prices $C^{\tiny{Model}}(K_i),\  i=1, \dots, N$ and the given option market price $C_i,\  i=1, \dots, N$, we used the {\it Mean Squared Error}, MSE,
$$
MSE(Model)= \frac{1}{N}\sum_{i=1}^{N}(C^{{{Model}}}(K_i)-C_i)^2. 
$$
 Clearly, it is expected that within the scope of the SV model described in ({\ref{40}), the well-calibrated Heston's model will results with a smaller Mean Squared Error as compared to the Black-Scholes model, so that $MSE(HS)\leq MSE(BS)$.  However, as we will see below for the available ETF data, pricing the options by a well-calibrated Generalized Gamma (GG) model (\ref{18}) also resulted with a smaller  MSE. In fact, in all four cases, $MSE(GG)\leq MSE(BS)$.  To demonstrate this, we have taken for each ETFs  the following steps, (conditional of course on the current spot's price $S$ and volatility $V_0$),
 
 \begin{itemize}
 \item Model Calibration,
 \begin{itemize}
  
 \item For a given model's parameter, $\vartheta=(\kappa, \theta, \eta, \rho)$ in ({40}), we use the {\tt callHestoncf} function of the NMOF package of R to calculate the Heston's  model option prices $C_i^{\small HS}$ for each $K_i$.
 
 \item To calibrate the Heston SV model, we used the {\tt optim($\cdot$)} function of R,  to minimize $MSE(HS)$  over the  model's parameter, 
  $\vartheta=(\kappa, \theta, \eta, \rho)$.

\item For a given $(\alpha, \nu)$ with $\nu=\sigma\sqrt{t}$, we use (\ref{18}) to calculate the Generalized Gamma model option prices $C_i^{\small GG}$ for each $K_i$.

\item To calibrate the GG model, we used the {\tt optim($\cdot$)} function of R,  to minimize $MSE(GG)$  over the  model's parameters, $(\alpha, \nu)$.  

\item For a given $\nu$ (where $\nu=\sigma\sqrt{t}$), we use (\ref{1})-(\ref{2}) to calculate the Black-Scholes model option prices $C_i^{\small BS}$ for each $K_i$.
 
 \item To calibrate the BS model, we used the {\tt optimize($\cdot$)} function of R,  to minimize $MSE(BS)$  over the single model's parameter
  $\nu$, (namely $\sigma$). 
 \end{itemize}
 \item Validation
 
    \begin{itemize}
    \item Using the calibrated Heston's parameters, $\hat \vartheta$ we drew,  utilizing a discretized version of Heston's stochastic volatility process (\ref{40}), a large number ($M=30,000$) of Monte-Carlo simulation,  observations on $(S_T, V_T)$ to 
obtain the simulated rendition of the Heston's RND of $S_t$, (conditional on $S$ and $V_0$, with $t=T-\tau$). 

\item Using the calibrated Heston's parameters, $\hat \vartheta$ in (\ref{41}) we obtain the calculated rendition of the Heston's theoretical RND of $S_t$, (conditional on $S$ and $V_0$, with $t=T-\tau$), directly from the characteristics function of $P_2$ (see Appendix). 

\item Compare all three calibrated risk-neutral distributions of the standardized spot's price (the rescaled spot priced, $S^*_t=S_t/\mu$, where $\mu=Se^{rt}$)  as obtained under the Black-Scholes (BS), Generalized Gamma (GG) and  Heston's (1993)  option pricing models (HS). 

\end{itemize}
 
 \end{itemize}
 
  \subsection{Calculating the implied RND under the volatility skew} 
  
 As we mentioned earlier, the data on the October 15, 2021 option series of the {\tt SPY, IWM} and {\tt QQQ} were retrieved as of the closing of trading on Friday August 13, 2021 with 63 days to expiration, so that $t=63/365$ and the prevailing (risk-free) interest rate is $r=0.0016$.  This will be common values for this  three highly liquid ETFs. The  October 15, 2021 option series of the{\tt TLT} was retrieved on August 18, 2021 with 57 days to expiration, so that $t=57/365$ for that ETF. However, we begin our exposition with the details of the largest (volume-wise) of them,  namely the {\tt SPY}.  The cases of the {\tt IWM},  {\tt QQQ} and {\tt TLT} will be treated similarly below. 
  
  On that day, the closing price of the {\tt SPY} was $S=445.92$ and the dividend it pays is at a rate of $\ell=0.0123$. We incorporate the dividend  in our calculations along the lines of Remark 1 in Boukai (2021).  The reported (BS-based) implied volatility was $IV=16.15\%$ which we will used as our initial value for $V_0$ and for $\sigma$.  This option series has $N=211$ pairs of strike-price $(K_i, \ C_i) $ which were all used to calibrate the Heston's SV model over the the  model's parameter , $\vartheta=(\kappa, \theta, \eta, \rho)$ with the initial values of $(15, (0.1)^2, 0.1, -0.65)$ and with $V_0=IV^2=(0.1615)^2$.  The results of the calibrated values are 
$$
\hat \vartheta =(15.03132587,\ 0.02793781, \ 2,\ -0.77469470).  
$$

\begin{figure}[t] 
  \begin{center}
  \includegraphics[width=4in,height=5in,keepaspectratio]{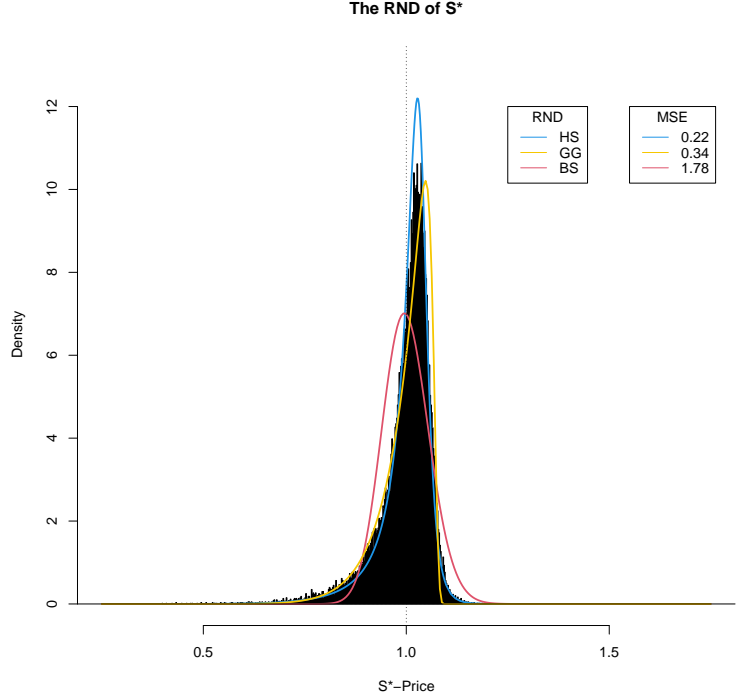}
  \small{\caption{\it{The {\tt SPY} case- the calculated HS, GG and BS implied RNDs along with the Monte-Carlo distribution of the Spot's price $S^*$, and the corresponding values of the MSE's}}}
  \end{center}
\end{figure}

This calibrated parameter, $\hat \vartheta $, was  then used to calculate, using Heston's characteristic function in (i.e. (\ref{40.5})), the option prices according to Heston's SV model (\ref{3}). This resulted with $MSE(HS)=0.2226429$. The calibrated (least squares estimate) value of $\sigma$ that minimizes $MSE(BS)$ is $\hat \sigma=0.137348$, so that $\hat \nu_{\tiny BS}=0.137348 \sqrt{t}=0.0570619$ to be used for the calculation of the $pdf$ of the ${\cal N}(-\nu^2/2, \, \nu^2)$ distribution  which leads to the BS formula in (\ref{1}) (see Example 3.1 in Boukai (2021) for more details). Next, we calibrated the General Gamma distribution according to the pricing model in (\ref{18}), with initial values of $\alpha=0.5$ and $\sigma=0.1615$ which resulted with calibrated value of $\hat \alpha=0.1554312$ and $\hat \sigma= 0.1483843$ and an $MSE(GG)=0.339441$ (see Table 3 below). Clearly, in this case of the SPY, the MSE of the GG pricing model is substantially smaller than the MSE of the Black-Scholes model and is similar to that of the Heston's SV pricing model. Indeed, the MSE of the BS model is over 500\% as large as those of the GG and the HS models.

  To compare the actual distributions, as calculated under each of these three pricing models, we present in Figure 2 the RNDs of the three implied distributions, calculated base on their respective calibrated parameter values.  As an added validation, we plotted these three density curves against 
 the histogram of the Monte-Carlo simulation of the standardized SPY prices using a discretized version of the pricing model in (\ref{40}) (using the calibrated Heston's parameters with a seed=452361).  This figure clearly demonstrates the `inaptness' of the standard  BS formula (\ref{1}) and hence the log-normal distribution,  for modeling option prices in cases which involve negatively skewed price distributions. In fact, the calculated values of the Kurtosis  and Skewness measures of each these distributions (see Table 1)  is also indicative of the noted lack-of-fit of the BS model in these cases and the apparent close agreement of the GG distribution to the exact  risk-neutral distribution of the Heston's model and that of the simulated price data.

 \begin{table}[h]
\begin{center}{\small 
\caption{\it {Calculated (excess) Kurtosis and Skewness measures for the three distributions depicted in Figure 1 for the {\tt SPY} option  data.}}
\begin{tabular}{ccccc}
\hline
&  Measure  & HS  & GG & BS \  \\ \hline
& Kurtosis	 &	7.302674 &	3.536461	& 0.05234164\  \\ 
& Skewness &	-2.050771  & -1.580122 & 0.1715114\  \\ 
 \hline
\end{tabular}
}
\end{center}
\end{table}
 \vskip -0.25truein
 Also of interest is the impact of this model's misspecification on the calculated delta values associated with the option series. It is a standard practice of the retail brokerage houses to provide, along with the market prices for the option chain, also the BS-base calculated delta for each strike (using some ATM implied volatility value). For example,   for the ATM strike of $K=445$ the quoted delta is $\Delta^*=0.497$ with quoted $IV$ of $0.1489$ whereas under the BS model we calibrated here with $\hat \sigma=0.137348$, we obtained $\Delta_{\small{BS}}=0.506$. However, accounting for  stochastic volatility in the pricing model, we calculate for this same strike, $K=445$, $\Delta_{\small{HS}}=0.663$ by the (better fitting) Heston SV model, and  $\Delta_{\small{GG}}=0.638$, by its close proxy, the GG model. Thus in this case, the BS modeling at the ATM strikes will result with grossly understated delta values (of nearly 25.0\%).  Without doubt, the impact of this model's misspecification would have profound hedging implications for the retail trader.  To fully appreciate ithe extent of this impact, we present in Figure 3, the values of the Delta function (\ref{8})  as was calculated for the HS model (using $P1$ and (\ref{40})), for the GG model (using (\ref{16})) and for the BS model (using $\Phi(d_1)$ from (\ref{1})), along with quoted delta values for the {\tt SPY} chain.

\begin{figure}[h] 
  \begin{center}
  \includegraphics[width=4in,height=5in,keepaspectratio]{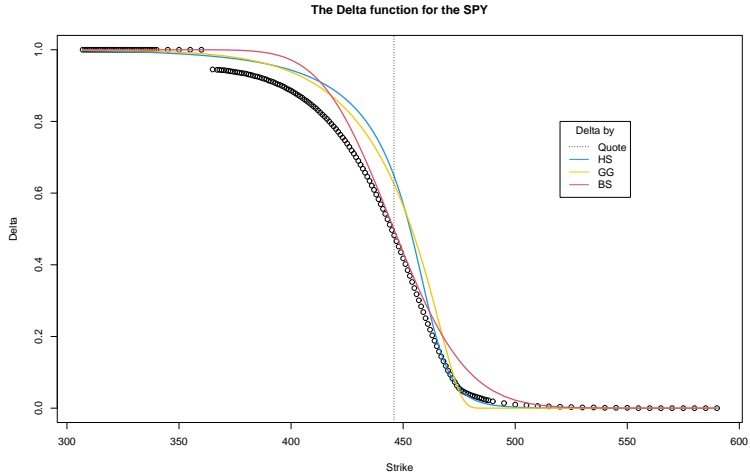}
 \small{ \caption{\it{The {\tt SPY} case- the calculated delta functions  under each of the pricing models, HS, GG and BS, along with the quoted delta per each strike $K$ in the October 15, 2021 option series.}}}
\end{center}
\end{figure}

\begin{figure}[h]
\centering
\begin{subfigure}{.6\textwidth}
  \centering
  \includegraphics[width=1\linewidth]{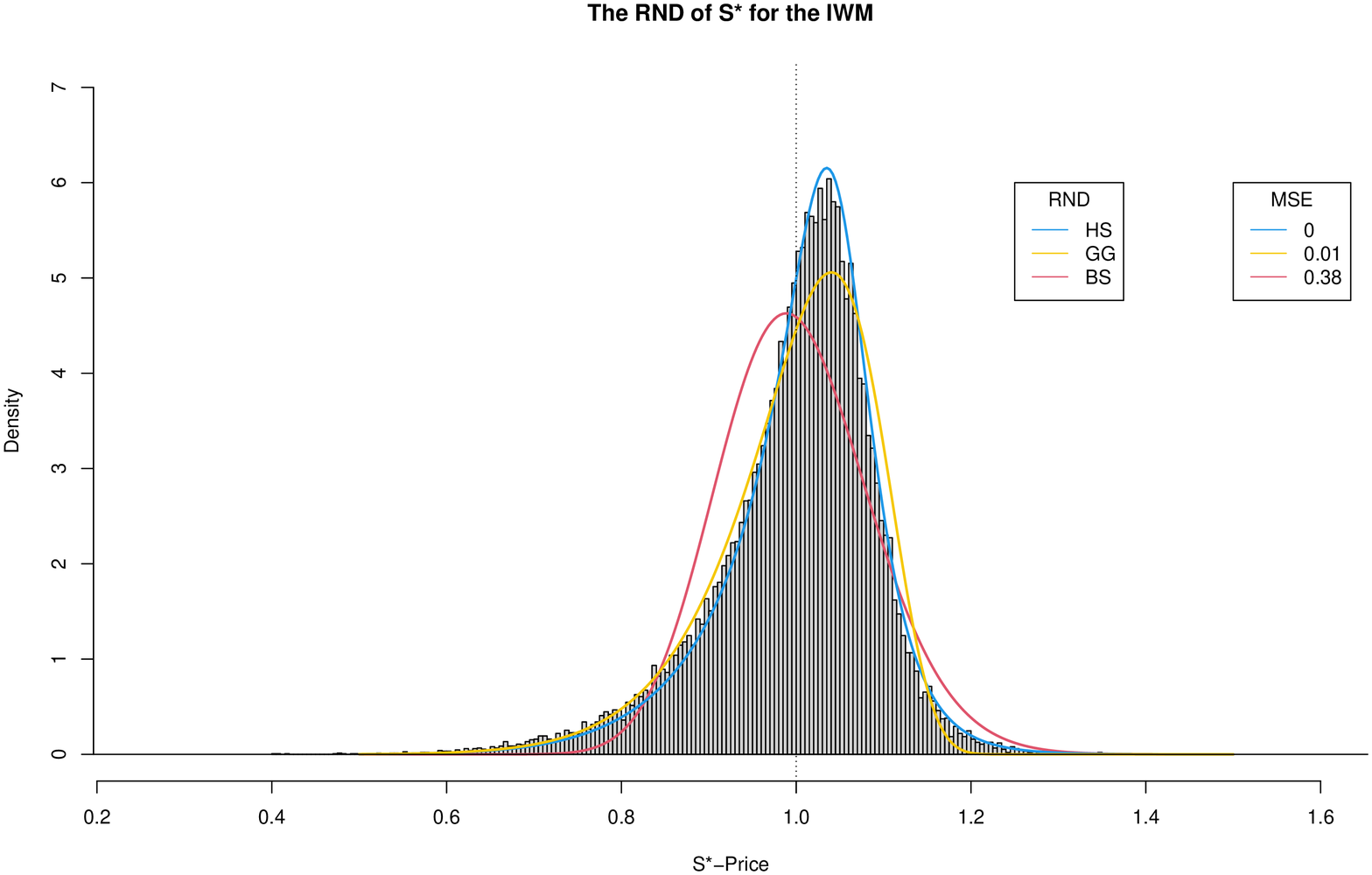}
  \caption{Implied RNDs  }
  \label{fig:sub1}
\end{subfigure}%
\begin{subfigure}{.6\textwidth}
  \centering
  \includegraphics[width=1\linewidth]{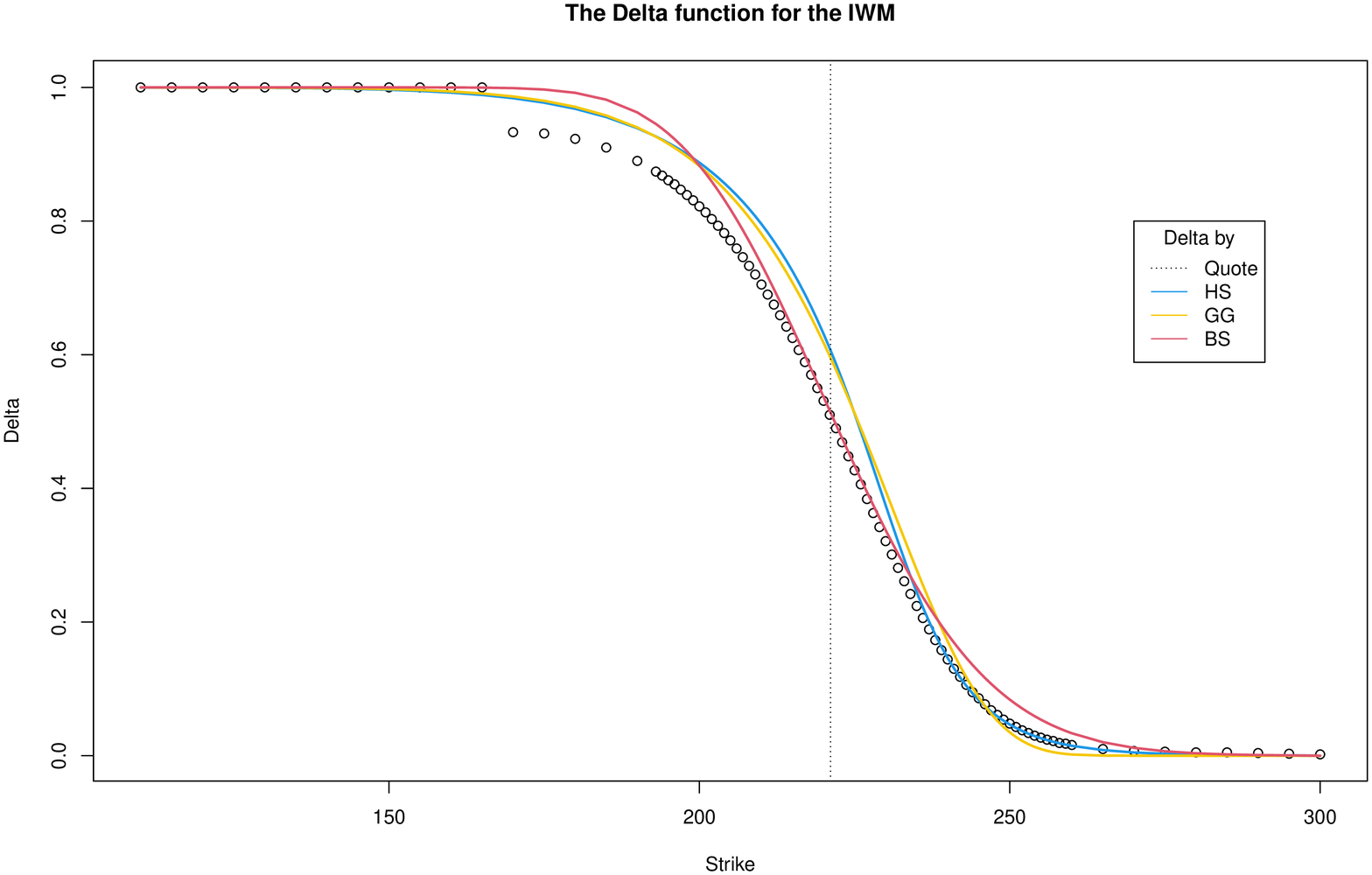} 
  \caption{Delta function }
  \label{fig:sub2}
\end{subfigure}{\small
\caption{\it{The {\tt IWM} case- (a) the HS, GG and BS implied RNDs along with the Monte-Carlo distribution of the Spot's price $S^*$, and (b) the corresponding delta functions  along with the quoted delta per each strike $K$ in the option series.}}}
\end{figure}

\begin{figure}[h]
\centering
\begin{subfigure}{.6\textwidth}
  \centering
  \includegraphics[width=1\linewidth]{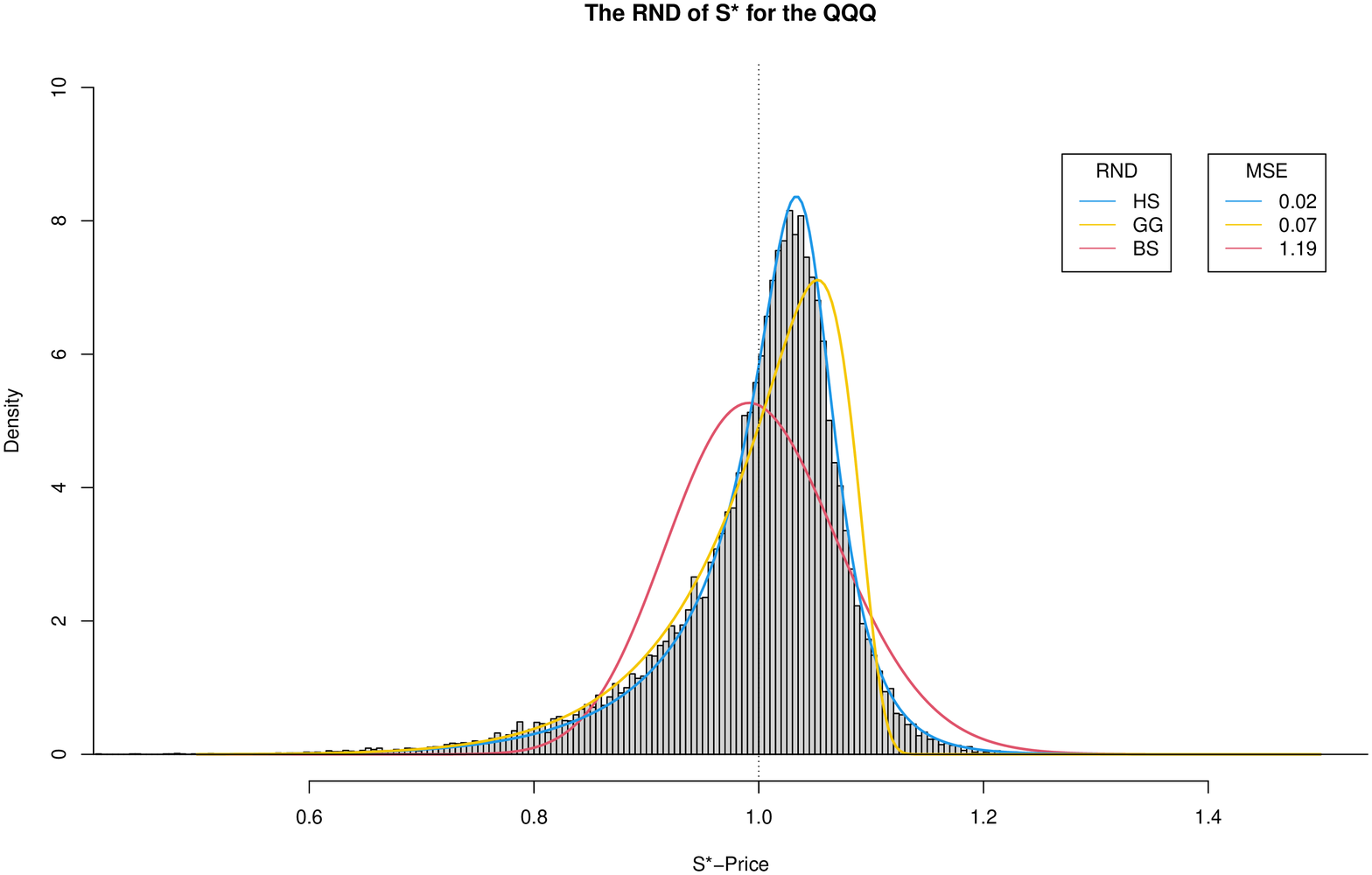}
  \caption{Implied RNDs}
  \label{fig:sub1}
\end{subfigure}%
\begin{subfigure}{.6\textwidth}
  \centering
  \includegraphics[width=1\linewidth]{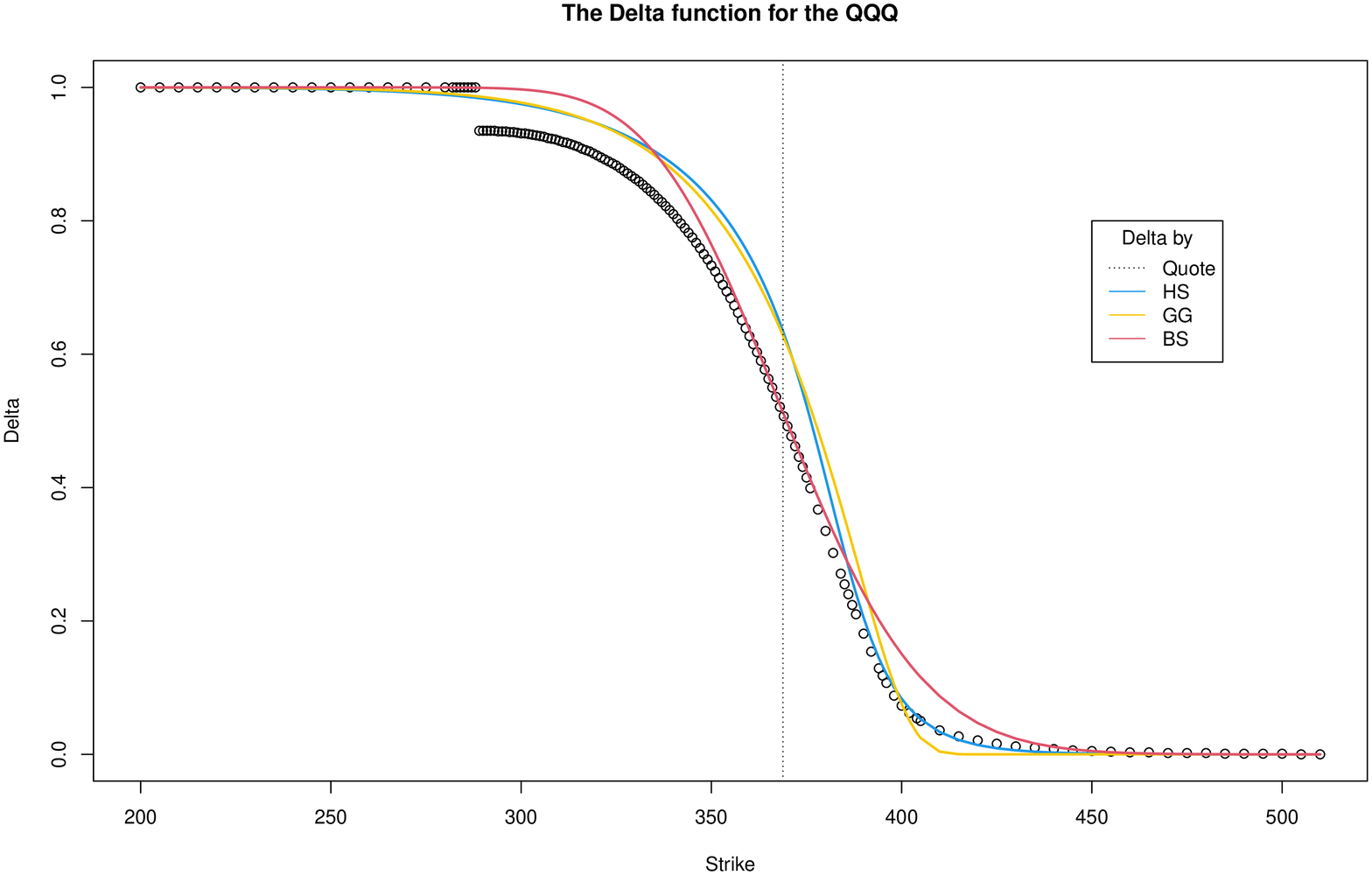}
  \caption{Delta functions }
  \label{fig:sub2}
\end{subfigure}
\caption{\small{\it{The {\tt QQQ} case- (a) the HS, GG and BS implied RNDs along with the Monte-Carlo distribution of the Spot's price $S^*$, and (b) the corresponding delta functions  along with the quoted delta per each strike $K$ in the option series.}}}
\label{fig:test}
\end{figure}

 Needless to say, the noted understatement of the quoted (BS -based) delta values as compared to those derived from the SV model also impact the trading strategies. For example, a trader that would sell a 25-Delta strangle based on the quoted values, will sell the $k_1=424$ put for \$5.215 and the $k_2=460$ call for \$2.685, collecting a total of \$7.90 for it, which amount to 21.9\% of the spread between the strikes (for a discussion of this ratio, see Boukai (2020)). On the other hand, if the trader would have priced the  25-Delta strangle according to the GG model (which accounts for the skew), she will sell  the $k_1=435$ put for \$7.205 and the $k_2=466$ call for \$1.435, collecting a total of \$8.64 for it, which amount to 27.9\% of the spread between the strikes. Clearly collecting a higher premium  for the same 25-delta strangle.

 The situation with the other two  market-index ETF's,  {\tt IWM} and {\tt QQQ},  is very similar to the one describing the {\tt SPY}--see the corresponding depiction of their volatility `smiles' in Figure 1.  Following a similar calibration and validation approach, we present (implied)  the price distributions derived from the {\tt IWM}  option data Figure 4 (a)  and the {\tt QQQ} option data Figure 5 (a).  The  calculated values of the corresponding delta functions are displayed in Figure 4 (b) and Figure 5 (b).  
 Further, to serve as contrasting illustration, we present in Figure 6  the three implied price distributions  derived from the  {\tt TLT} ETF option series, along with the corresponding calculated delta functions for that ETF.  The situation with the {\tt TLT} ETF  is clearly different, as compared to the three market-index ETFs ({\tt SPY,  IWM} and {\tt QQQ}) which exhibit pronounced skew of their volatility `smile'.   In the case of the {\tt TLT})  ETF, with a relatively intact volatility `smile' (see Figure 1) the implied RNDs are relatively symmetric and three option pricing models (HS, GG and BS) yield very similar results.   In Table 2 we provide a summary of the {\it goodness-of-fit}  of each of the pricing models as measured by the respective MSE for each of the four ETFs.   A corresponding comparison of the ATM delta calculations under each of the option price models is presented in Table 3.   In Table 3 we provide a summary of the {\it goodness-of-fit} as measured by the respective MSE for each of the ETFs.  Some of the technical details are provided in Section 4.1 below.

 \begin{table}[h]
\begin{center}{\small 
\caption{The {\it goodness-of-fit} as measured the respective MSE for each of the four ETFs .}
\begin{tabular}{ccccc}
\hline
&  ETF  & HS  & GG & BS \  \\ \hline
& {\tt SPY} 	 &	 0.2226429 & 0.339441  &  1.781981\  \\ 
& {\tt IWM} &	0.001900968 & 0.01419628 & 0.3750478 \  \\ 
& {\tt QQQ} &	0.02418013 & 0.06561134 & 1.193867\  \\ 
& {\tt TLT} &	0.03423748 & 0.04618725 & 0.04341321\  \\ 
 \hline

\end{tabular}
}
\end{center}
\end{table}

 \begin{table}[h]
\begin{center}{\small 
\caption{\it{Comparison of the the quoted ATM delta $\Delta^*$ of the four market ETF to those calculated under each of the three option pricing models. }}
\begin{tabular}{ccccccccc}
\hline
&  ETF  &  $S$ & ATM $K$ & $\Delta^*$ & $\Delta_{\small{BS}}$   & $\Delta_{\small{GG}}$  & $\Delta_{\small{HS}}$ &  \  \\ \hline
& {\tt SPY} 	 & 445.92 &  445	& 0.497 &   0.506 & 0.638  & 0.663  \  \\ 
& {\tt IWM} &	 221.13  & 221  & 0.510 &  0.516 & 0.598  & 0.610   \  \\ 
& {\tt QQQ} &  368.82 & 369	& 0.507 & 0.503 & 0.625  &  0.632 \  \\ 
& {\tt TLT} &  149.35  &150	& 0.511 &  0.453 & 0.477   &  0.467 \  \\ 
 \hline

\end{tabular}
}
\end{center}
\end{table}

  \section{Summary and discussion}
 
As was illustrated in all the above examples, the Heston (1993) option pricing model (as in (\ref{40}) and (\ref{3})) which accounts for the presences of stochastic volatility, produces as expected, the best results overall as compared to the Black-Scholes option pricing model (\ref{1}) with its presumed constant volatility.
If available, a well-calibrated Heston's model, will always result in a better fit to realistic market option data (indeed, resulting with $MSE(HS)< MSE(BS)$) and would be the default modeling choice for the practitioner. 
Unfortunately however, the complex calculations and calibration process of the Heston's option pricing model (see Appendix) renders it inaccessible to many of the retail option traders. 
In comparison, the calculations and calibration of the option pricing model under the Generalized Gamma distribution as RND are substantially simpler and straightforward (and could potentially be accomplished within and Excel spreadsheet). As was demonstrated earlier, the GG model is significantly more accurate than the Black-Scholes model for the pricing of the options in a skewed stochastic volatility environments as those exhibited (\t{at present times}) by the three markets ETFs, {{\tt SPY, IWM}} and {{\tt QQQ}}. 
In fact, in situations which imply negatively skewed price-distributions as RND, the Black-Scholes pricing model, and hence the log-normal distribution as RND, will surely be inferior to the Generalized Gamma distribution, and surely to Heston's SV pricing model in fitting realistic option market data. In such situations one would realize $MSE(GG)< MSE(BS)$ and would want to adopt the GG distribution as RND for the underlying pricing model. In contrast, in situations such as the one exhibited by the {\tt TLT} ETF, one would realize $MSE(GG)\approx MSE(BS)$, as all three option pricing models (including Heston's) produce similar results. 
Although not expressly covered by the examples we analyzed here, we have grounds to believe that the same conclusion could be arrived upon using the Inverse Generalized Gamma (see Section 2.2) as an RND in situations involving positively skewed (implied) RND in the option pricing model. In all, both of these versions of the Generalized Gamma distribution could serve as useful proxies to the exact Heston's RND and hence produce superior results to those obtained by the Black-Scholes model in an environment involving stochastic volatility. Thus given the market option data, one could simply calculate $MSE(BS), MSE(GG)$ (and if needed also $MSE(IGG)$) and adopt the option pricing model which  produces the better fit.

\subsection{Some technical notes}
\begin{itemize}

\item The the October 15, 2021 option series data files {\tt SPY\_63.csv, IWM\_63.csv}, and  {\tt QQQ\_63.csv} as were obtained on the EOD of August 13, 2021 and that of {\tt TLT\_57.csv} obtained at the EOD of August 18, 2021, are available from the author upon request.  Their basic summary information is provided in Table 4 below.

  \begin{table}[h]
\begin{center}{\small 
\caption{\it{Summary information of the four ETFs .}}
\begin{tabular}{ccccccccc}
\hline
&  ETF  &  $S$ & DTE  & N& Quoted IV   &    Div.  Rate   \  \\ \hline
& {\tt SPY} 	 & 445.92 &  63	& 211&  16.15\%  & 1.23\%    \  \\ 
& {\tt IWM} &	 221.13  & 63 &93 &  24.30\%  & 0.63\%   \  \\ 
& {\tt QQQ} &  368.82 & 63	& 160  & 18.13\%  &0.43\%   \  \\ 
& {\tt TLT} &  149.35  &57	& 66&  15.71\% & 1.46\%  \  \\ 
 \hline

\end{tabular}
}
\end{center}
\end{table}

\item The R function {\tt dgamma} and {\tt pgamma} were used to calculate the $pdf$ and $cdf$ in (\ref{10}) and hence in the calculation of (\ref{18}). 

\item The {\tt cfHeston} and {\tt callHestoncf} functions of the NMOF package of R, were used in the calculation of (\ref{40.5}) and (\ref{3}).
\item A modification of the {\tt callHestoncf} function of the NMOF package of R was used to calculate (\ref{41}).
 
 \item The {\tt optim} and {\tt optimize} functions of R were used in the calibration of the three models (HS, GG and BS)  for the available option data.

\item The initial and the calibrated values of $\vartheta=(\kappa, \theta, \eta, \rho)$ of the Heston's model were:
\begin{itemize} 
\item[{\tt SPY}:]  $(15, (0.1)^2, 0.1, -0.65)$\ and \ $(15.03132587,\ 0.02793781,\ 2,\ -0.77469470)$.
\item[{\tt IWM}:]  $(5,  (0.1)^2, 0.6, 0)$\ and \  $(4.97834286,  0.04032166, 1.09837930, -0.59905916)$.
\item[{\tt QQQ}:] $(3.5, (0.2)^2, 0.5, -0.5)$ \  and \ $(3.47635183, 0.06382197, 1.13505528,   -0.69137767)$. 
\item[{\tt TLT}:] $(3, (0.1)^2, 0.1, 0.1)$ \  and \ $(2.99997881, 0.01459405, 0.10011507,   0.10007980)$.

\end{itemize}

\item  For the Monte-Carlo simulation of (\ref{40}) we employed  the (reflective version of) Milstein’s (1975) discretization
scheme (see also Gatheral (2006)) with seeds =4569 ({\tt QQQ}), =777999 ({\tt IWM}) and  =452361 ({\tt SPY}) and  = 121290 ({\tt TLT}).

  \end{itemize}
  \bigskip

\begin{figure}[h]
\centering
\begin{subfigure}{.6\textwidth}
  \centering
  \includegraphics[width=1\linewidth]{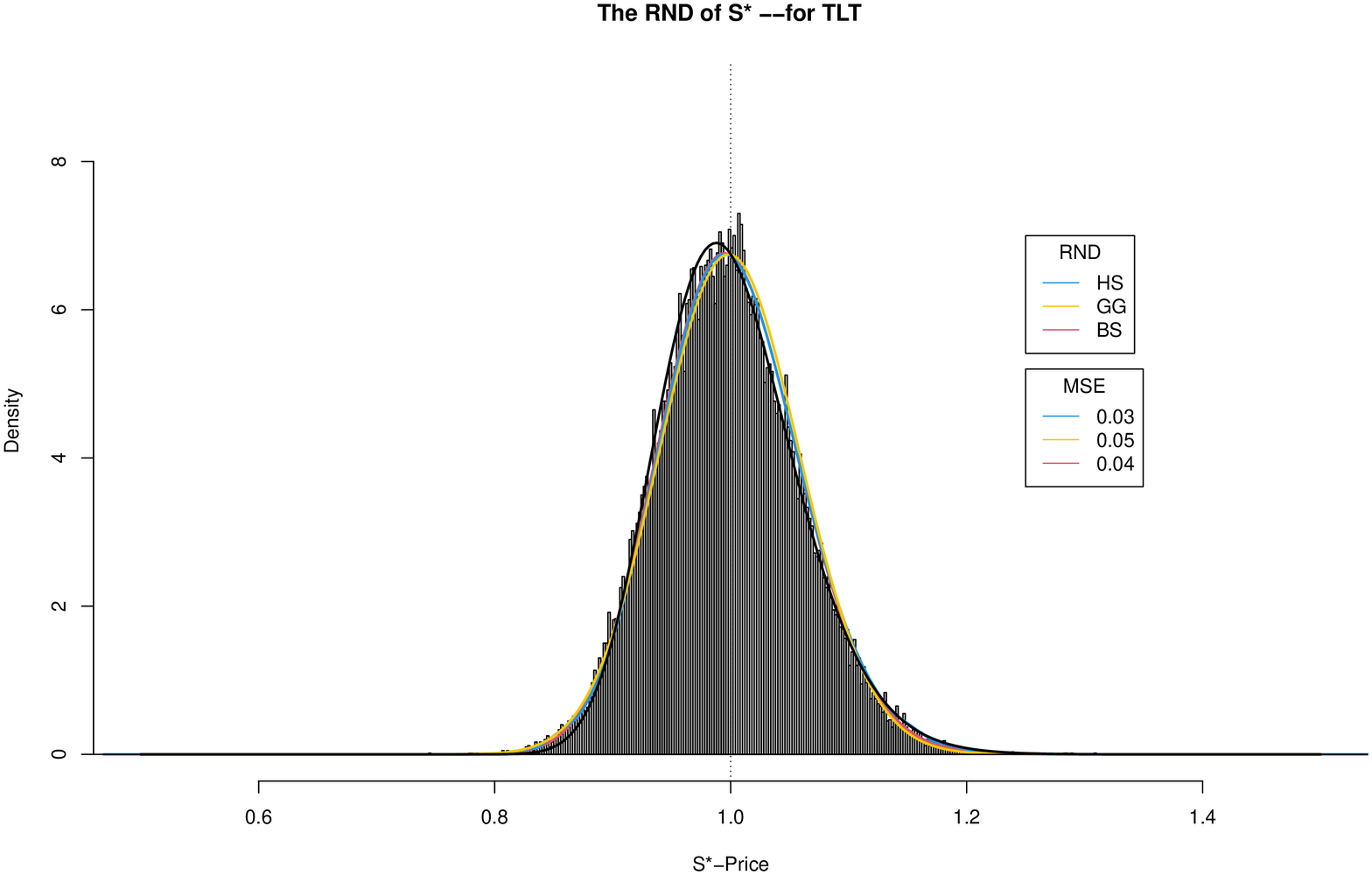}
  \caption{Implied RNDs}
  \label{fig:sub1}
\end{subfigure}%
\begin{subfigure}{.6\textwidth}
  \centering
  \includegraphics[width=1\linewidth]{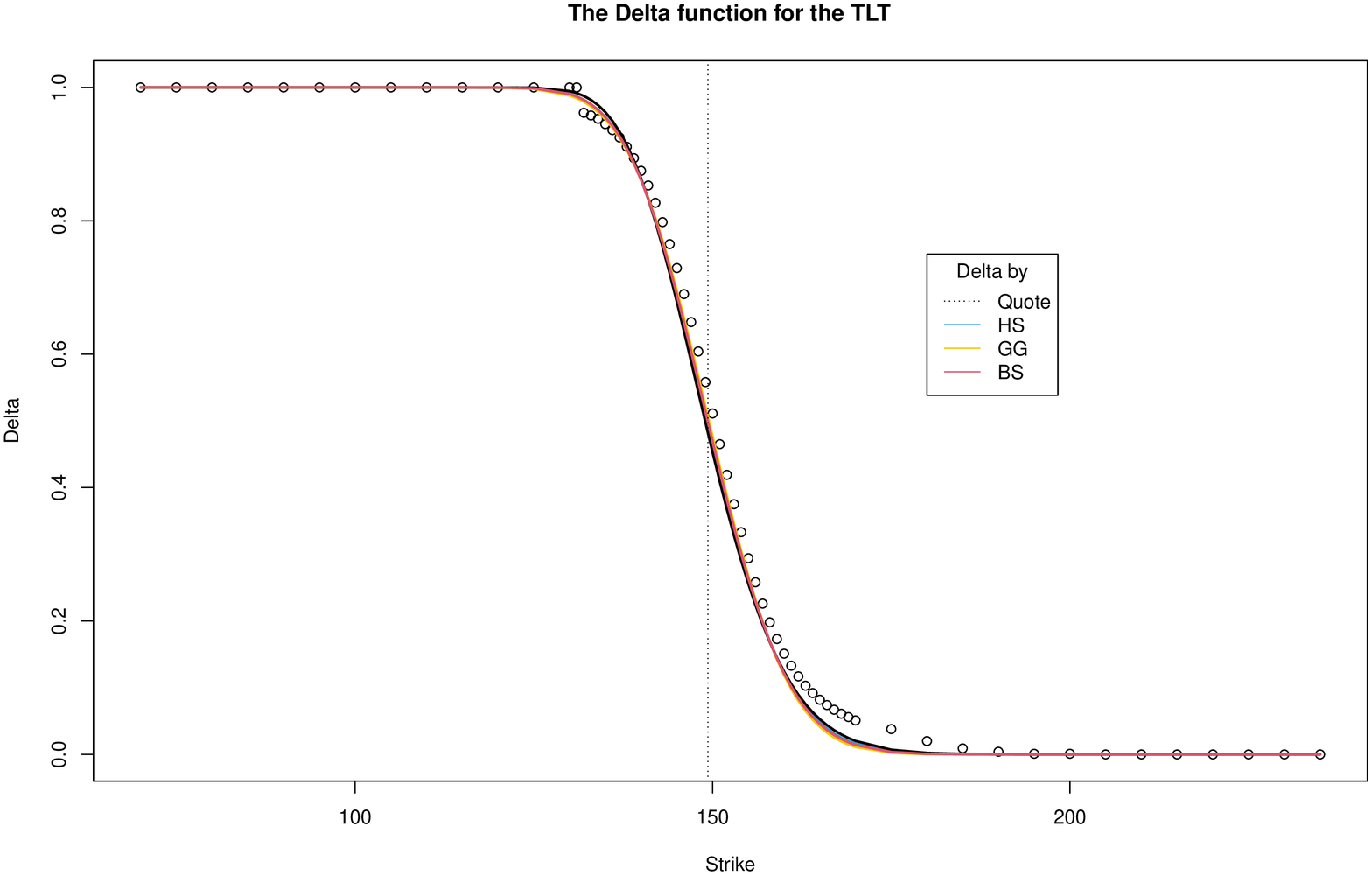}
  \caption{Delta function}
  \label{fig:sub2}
\end{subfigure}\small{
\caption{\it{The {\tt TLT} case- (a) the HS, GG and BS implied RNDs along with the Monte-Carlo distribution of the Spot's price $S^*$, and (b) the corresponding delta functions  along with the quoted delta per each strike $K$ in the option series.}}}
\label{fig:test}
\end{figure}

\vfill\eject

\section{Appendix}

Heston (1993) considered the  stochastic volatility model describing the price-volatility dynamics (of ${\cal{S}}=\{S_t, \, t\geq 0\}$ and ${\cal{V}}=\{V_t, \, t\geq 0\}$) as described via a system of stochastic deferential equations given by,
\be\label{40}
\begin{aligned}
dS_t= & rS_tdt +\sqrt{V_t}S_t dW_{1,t}\\
dV_t= & \kappa(\theta-V_t)+\eta\sqrt{V_t}dW_{2,t}, 
\end{aligned}
\ee
where $r$ is the risk-free interest rate,  $\kappa, \ \theta$ and $\eta$ are some constants (as discussed in Section 1) and where  $W_{1}=\{W_{1,t}, \, t\geq 0\}$ and  $W_{2}=\{W_{2,t}, \, t\geq 0\}$ are two Brownian motion processes under (under the risk neutral probability $\Q$)  with $d(W_{1}W_{2})=\rho dt$ for some $\rho^2\in(0,1)$. 
Heston (1993)  offered $C_S(K)$ in (\ref{3})  as the solution to the option valuation under the above SDE  and provided  (semi) closed form expressions to the probabilities $P_1$ and $P_2$ that comprise it.

These closed form expressions are given  for $j=1, 2$ by,
\be\label{40.5}
P_j=\frac{1}{2}+\frac{1}{\pi}\int_{0}^\infty{\cal R}\text{e}\left[ \frac{e^{-i\omega k}\psi_j(\omega, t, v, x)}{i\omega}\right]d\omega,
\ee
where  with $x:=\log{(S)}$, $k:=\log{(K)}$, $ b_{1}=\kappa - \rho \eta, \ \ b_{2}= \kappa $ and $\psi_j(\cdot)$ is the  characteristics function 
$$
\psi_j(\omega, t, v, x):=\int_{-\infty}^\infty e^{i \omega s}p_j(s)ds\equiv e^{B_j(\omega, t)+D_j(\omega, t)v+i \omega x+ i\omega \,  rt}. 
$$
Here $p_j(\cdot)$ is the $pdf$ of $s_{T}=\log(S_T)$ corresponding to the probability $P_j, \ j=1, 2$ and 
$$
B_j(\omega, t)= \frac{\kappa \theta}{\eta^{2}}\{(b_{j} +d_j- i \omega \rho \eta)t -2\log(\frac{1- g_j e^{d_jt}}{1-g_j})\}
$$
$$
D_j(\omega, t)= \frac{b_{j} +d_j -i\omega \rho \eta}{\eta^{2}}(\frac{1-e^{d_jt}}{1-g_je^{d_jt}})
$$
$$
g_j=\frac{b_{j}-i\omega\, \rho \eta  +d_j}{b_{j}-i\omega\rho \eta  -d_j}
$$
$$
d_j= \sqrt{( i\omega \rho \eta - b_{j})^{2}- \eta^{2}(2 i\omega u_{j} -\omega^{2})}. 
$$
Note that by a standard application of the Fourier transform obtains  (see for example Schmelzle (2010)) that the $pdf$ $p_j(\cdot)$ of $s_{T}=\log(S_T)$, can be computed,  for any $s\in {\Bbb R}$, as
\be\label{41}
p_j(s)=\frac{1}{\pi}\int_{0}^\infty{\cal R}\text{e}\left[ {e^{-i\omega s}\psi_j(\omega, t, v, x)}\right]d\omega. 
\ee
Regardless of their complexities, efficient numerical routines such as the  
{\tt cfHeston} and {\tt callHestoncf} functions of the NMOF package of R, are readily available nowadays to accurately compute the values of $\psi_j$ and hence of $P_j$ and the call option values $C_S(K)$ in (\ref{3}), for given $t, s$ and $v$ and any choice of $\vartheta=(\kappa, \theta, \eta, \rho)$.


\begin{thebibliography}{999}

\bibitem{} Albrecher, H., Mayer, P., Schoutens,  W. and Tistaer, J., (2007).
\newblock The Little Heston Trap. 
\newblock\emph{The Wilmott Magazine}, 83–92.

\bibitem{} Alfonsi A., (2010)
\newblock High order discretization schemes for the CIR process: Application to affine term structure and Heston models. 
\emph{Math. Comp.} 79(269),  209–237.

\bibitem{} Andersen L., (2008)
\newblock Simple and efficient simulation of the Heston stochastic volatility model. 
\newblock \emph{J. Comput. Finance},  11(3), 2008, 1–42. 

\bibitem{} Black F., and Scholes M., (1973).
\newblock The pricing of options and corporate liabilities.
\newblock \emph{The Journal of Political Economy}, 637-654.

\bibitem{} Bakshi G., Cao C. and Chen Z., (1997) .
\newblock Empirical Performance of Alternative Option Pricing Models.
\newblock \emph{The Journal of Finance}, Vol. LII, No. 5, 2003-2049.

\bibitem{} Boukai B. (2020).
\newblock How Much is your Strangle Worth? On the Relative Value of the delta-Symmetric Strangle under the Black-Scholes Model
\newblock \emph{Applied Economics and Finance}, Vol. 7, No. 4; July 2020 \url{https://doi.org/10.11114/aef.v7i4.4887}

\bibitem{} Boukai B. (2021).
\newblock On the RND under Heston's Stochastic Volatility Model.
\newblock Available at \emph{SSRN}\url{https://ssrn.com/abstract=3763494}

\bibitem{} Cox J.C. and Ross S., (1976)
\newblock The valuation of options for alternative stochastic processes.
\newblock \emph{J. Fin. Econ.} 3:145-66.

\bibitem{}  Feller W., (1951)
\newblock Two singular diffusion problems. 
\newblock \emph{Ann. Math.}  54(1),  173–182. 

\bibitem{} Figlewski S., (2010). 
\newblock ``Estimating the Implied Risk Neutral Density for the U.S. Market Portfolio'',  in \emph{Volatility and Time Series Econometrics: Essays in Honor of Robert F. Engle, (eds. Tim Bollerslev, Jeffrey Russell and Mark Watson)} 
\newblock Oxford University Press, Oxford, U.K. 

\bibitem{} Figlewski S.,  (2018). 
\newblock{} Risk Neutral Densities: A Review.
\newblock Available at \emph{SSRN}\url{ http://ssrn.com/abstract=3120028}.
 
\bibitem{} Gatheral, J., (2006). 
\newblock \emph{The Volatility Surface,}
\newblock  John Wiley and Sons, NJ.

\bibitem{} Grith M. and Krätschmer V.,  (2012) 
\newblock ``Parametric Estimation of Risk Neutral Density Functions'', in: \emph{Ed: Duan JC., Härdle W., Gentle J. (eds) Handbook of Computational Finance}
\newblock Springer Handbooks of Computational Statistics. Springer, Berlin, Heidelberg.

\bibitem{} Heston S.L., (1993)
\newblock A closed-form solution for options with stochastic volatility with applications to bond and currency options. 
\newblock\emph{Rev. Financ. Stud.}  6(2),  327–343. 

\bibitem{} Jackwerth, J. C., (2004).
\newblock \emph{Option-Implied Risk-Neutral Distributions and Risk Aversion}
\newblock Research Foundation of AIMR, Charlotteville, NC

\bibitem{} Jackwerth, J. C. and Rubinstein, M., (1996).
 \newblock Recovering Probability Distributions from Option Prices.
 \newblock \emph{ The Journal of Finance,}  51, no. 5: 1611-631. 

\bibitem{} Jiang, L. (2005).
\newblock \emph{Mathematical Modeling and Methods of Option Pricing,}
\newblock Translated from Chinese by Li. C,
\newblock   World Scientific, Singapore.

\bibitem{} Kiche J.,  Ngesa O. and Orwa, G. (2019)
\newblock On Generalized Gamma Distribution and Its Application to Survival Data
\newblock\emph{ International Journal of Statistics and Probability; Vol. 8, No. 5, \url{: https://doi.org/10.5539/ijsp.v8n5p85
}} 

\bibitem{} Lemaire, V.,  Montes, T.  and Pagès, G.,  (2020).
\newblock Stationary Heston model: Calibration and Pricing of exotics using Product Recursive Quantization. 
\newblock Available at \emph{arXiv} [q-fin.MF]: \url{arXiv:2001.03101v2 }.

\bibitem{} Merton, R., (1973).
\newblock Theory of rational option pricing.
\newblock \textit{The Bell Journal of Economics and Management Science}, 141-183.

\bibitem{} Mil’shtein, G. N. (1975). 
\newblock Approximate Integration of Stochastic Differential Equations. 
\newblock \emph{Theory of Probability \& Its Applications}, 19 (3), 557–562.

\bibitem{} Mizrach, B. (2010). 
\newblock ``Estimating Implied Probabilities from Option Prices and the Underlying'' in \emph{ Handbook of Quantitative Finance and Risk Management (C.-F. Lee A. Lee and J.  Lee (eds.))}.
\newblock Springer Science Business Media. 

\bibitem{} Mrázek, M. and Pospíšil, J. (2017).
\newblock Calibration and simulation of Heston model.
\newblock\emph{Open Mathematics| Vol.}  15(1), \url{https://doi.org/10.1515/math-2017-0058}.

\bibitem{} {{R Core Team}}, (2017). 
\newblock {\it R: A Language and Environment for Statistical Computing}.
\newblock {Vienna, Austria}, \url{https://www.R-project.org/}

\bibitem{} Stacy, EW.  (1962). 
\newblock A generalization of the gamma distribution. 
\newblock \emph{The Annals of Mathematical Statistics},  33, 1187-1192.


\bibitem{} Stein J. and Stein E., (1991).
\newblock Stock price distributions with stochastic volatility: An analytic approach. 
\newblock\emph{Rev. Financ. Stud.}  4(4),  727–752. 

\bibitem{} Schmelzle, M., (2010).
\newblock Option Pricing Formulae using Fourier Transform: Theory and Application.
\newblock \emph{Technical Report}, Available on line at \url{https://pfadintegral.com/articles/option-pricing-formulae-using-fourier-transform/}

\bibitem{} Savickas, R., (2002).
\newblock A simple option formula.
\newblock \emph{The Financial Review}, Vol 37, 207-226. 

\bibitem{} Savickas, R., (2005).
\newblock Evidence on delta hedging and implied volatilities for the Black-Scholes, gamma, and Weibull option pricing models.
\newblock \emph{The Journal of Financial Research}, Vol 18:2, 299-317.

\bibitem{} Thurai, M. and Bringi,  V. N. (2018)
\newblock Application of the Generalized Gamma Model to Represent the Full Rain Drop Size Distribution Spectra.
\newblock\emph{Journal of Applied Meteorology and Climatology}, Vol. 57, No. 5,  1197-1210.

\bibitem{}Wiggins, B. J., (1987).
\newblock Option values under stochastic volatility: Theory and empirical estimates.
\newblock \emph{Journal of Financial Economics}, Vol. 19(2), 351-372

\end{thebibliography}
\end{document}